\documentclass[reprint]{revtex4-1}

\usepackage{graphicx} 
\usepackage{amsmath}
\usepackage{ amssymb }

\usepackage{color}

\usepackage{float} 
\usepackage{wrapfig} 

\usepackage[makeroom]{cancel} 
\usepackage{comment}

\usepackage{lipsum} 

\usepackage{outlines}

\newcommand{\beq}{\begin{equation}}
	\newcommand{\eeq}{\end{equation}}
\newcommand{\vv}{\mathbf{v}}

\newcommand{\bvec}{\begin{pmatrix}}
	\newcommand{\evec}{\end{pmatrix}}
\newcommand{\lp}{\left(}
\newcommand{\rp}{\right)}
\newcommand{\pa}[2]{\frac{\partial #1}{\partial #2}}

\newcommand{\paf}[2]{\partial #1 / \partial #2}
\newcommand{\llangle}{\left \langle}
\newcommand{\rrangle}{\right \rangle}

\newcommand{\ve}[1]{\mathbf{#1}}

\newcommand{\tn}{\tilde{n}}

\newcommand{\tphi}{\tilde{\phi}}
\newcommand{\tE}{\tilde{E}}
\newcommand{\sgn}{\text{sgn}}

\renewcommand{\Im}[1]{\text{Im} #1}

\begin{document}



\title{Nonresonant Diffusion in Alpha Channeling}

\author{Ian E. Ochs and Nathaniel J. Fisch}
\affiliation{Department of Astrophysical Sciences, Princeton University, Princeton, New Jersey 08540, USA}


\date{\today}

\begin{abstract}

The gradient of fusion-born alpha particles that arises in a fusion reactor can be exploited to amplify waves, which cool the alpha particles while diffusively extracting them from the reactor. The corresponding extraction of the resonant alpha particle charge has been suggested as a mechanism to drive rotation. By deriving a coupled linear-quasilinear theory of alpha channeling, we show that, for a time-growing wave with a purely poloidal wavevector, a current in the nonresonant ions cancels the resonant alpha particle current, preventing the rotation drive but fueling the fusion reaction.

\end{abstract}

\maketitle

\textbf{Introduction:} 
A particle gyrating in a magnetic field with a velocity $v_\perp$ greater than the phase velocity $\omega_r / k_\perp$ of an electrostatic wave will become Landau-resonant at some point in its orbit, allowing for efficient wave-particle energy exchange.
Each time energy is exchanged with the wave, the particle gyrocenter position changes as well, leading to a wave-induced diffusion along a specified path in energy-gyrocenter space \cite{Fisch1992,fisch1992current}.
If the diffusion along this path on average pushes particles from higher to lower energy, the wave will amplify.
This effect is known as alpha channeling, so named because of its utility in simultaneously cooling and extracting alpha particles from the hot core of a fusion reactor, and channeling the resulting energy into useful wave power.

For alpha channeling in a slab geometry, the diffusion path slope in energy-gyrocenter space has the simple form $\paf{\ve{X}}{K} = \ve{k}\times \hat{b} / m_\alpha \omega \Omega_\alpha$, where $K$ is the perpendicular kinetic energy, $\ve{X}$ is the gyrocenter position, $m_\alpha$ and $\Omega_\alpha$ are the alpha particle mass and gyrofrequency, and $\omega$ and $\ve{k}$ are the wave frequency and wavenumber.
Thus, the condition for alpha channeling is:
\begin{align}
	\lp \pa{}{K} + \frac{\ve{k}\times \hat{b}}{m_\alpha \omega \Omega_\alpha} \cdot \pa{}{\ve{X}}\rp F_{\alpha 0} > 0, \label{eq:amplificationCondition}
\end{align}
where $F_{\alpha 0}$ is the zeroth-order distribution function in energy-gyrocenter space.

What remains unknown is whether or not the alpha particles carry net charge out of the plasma as a result of the wave-induced diffusion.
If charge is in fact carried out, then alpha channeling can be used drive $\ve{E} \times \ve{B}$ rotation in the plasma, providing an advantageous mechanism for shear rotation drive and centrifugal confinement in mirror fusion reactors \cite{fetterman2008alpha}.
Understanding whether such schemes are possible at all requires evaluation of the effect of the wave on the nonresonant particles, which has never been examined for alpha channeling.
Such reactions in the nonresonant particles are extremely important in enforcing momentum and energy conservation \cite{Kaufman1972a,kralltrivelpiece}, making theories that ignore them liable to error.

The reason the nonresonant response has proved elusive is that there is no existing linear theory of alpha channeling. 
Typically, a coupled linear wave and quasilinear particle system is necessary to calculate the nonresonant particle response.
The elusivity of the linear theory is related to the fact that Landau damping cannot be derived from the magnetized dispersion relation, a conundrum sometimes termed the Bernstein-Landau paradox \cite{Sukhorukov1997,charles2020bernsteinLandau}.
Derivation of the diffusion thus requires a nonlinear calculation, which allows for stochastic diffusion of the particle throughout phase space above a certain wave amplitude at which Landau-resonant particles dephase from the wave \cite{Karney1978,karney1979stochastic,sagdeev1973influence,dawson1983damping,Zaslavskii1986}.

In this Letter, we show that a linear-quasilinear system can be derived by assuming the wave-particle dephasing, which we accomplish by transforming the familiar \emph{unmagnetized} linear-quasilinear kinetic theory to gyrocenter coordinates.
To show that the system describes alpha channeling, we show that it recovers both the amplification condition (Eq.~(\ref{eq:amplificationCondition})) and the nonlinear diffusion coefficient \cite{karney1979stochastic} for channeling by lower hybrid (LH) waves.

Treating the channeling problem in this way positions us to answer the question of whether alpha channeling extracts charge from the fusion reactor.
We find that for the initial value problem in a slab geometry, where an electrostatic wave with purely poloidal wavenumber grows in time, the charge flux from the resonant alpha particles is canceled by an equal and opposite charge flux in the nonresonant particles, so that no reactor charging occurs.
We also determine which particles carry this nonresonant return current, in both single- and multi-ion-species plasmas.
For LH waves, the nonresonant return current is carried exclusively by fuel ions, so that alpha channeling has the added benefit of fueling the fusion reaction while extracting alpha particles.

\textbf{Linear theory:}
For any electrostatic wave, the dispersion relation obtained by linearizing and Fourier transforming Poisson's equation can be expressed as: 
\begin{align}
	0 &= 1 + \sum_s D_s; \quad D_s \equiv -\frac{4 \pi q_s }{k^2} \frac{\tn_s}{\tphi}, \label{eq:ESDispersionGeneral}
\end{align}
where $q_s$ and $n_s$ are the charge and density of species $s$, $\phi$ is the potential, and tildes denote Fourier transforms.
In the standard limit $|\omega_i / \omega_r| \ll 1$, this becomes:
\begin{align}
	0&=1 + \sum_s D_{r,s} \label{eq:realDisp}\\
	0&= \sum_s  \lp i \omega_i \pa{D_{r,s}}{\omega_r} + i D_{i,s} \rp \label{eq:imagDisp},
\end{align}
where $D_{r,s}$ and $D_{i,s}$ are the real and imaginary parts of the dispersion function Eq.~(\ref{eq:ESDispersionGeneral}) evaluated at real $\omega, k$.

To treat the alpha channeling initial value problem, we take the wavenumber $\ve{k} \parallel \hat{x}$, the magnetic field $\ve{B} \parallel \hat{z}$, and the gradient of the gyrocenter distribution function to be along $\hat{y}$.
Thus, $y$ corresponds to the ``radial'' direction, and $x$ to the ``poloidal'' direction.
For simplicity, we specialize to a LH wave, assuming cold fluid populations of magnetized electrons $e$ and unmagnetized ions $i$, and a hot, unmagnetized population of alpha particles $\alpha$.
Our dispersion components are thus given by: $D_{r,e} = \omega_{pe}^2/\Omega_e^2$, $D_{r,i} = -\omega_{pi}^2/\omega^2$, $D_{i,e} = D_{i,i} = 0$, and:
\begin{align}
	D_\alpha &= -\frac{\omega_{pi}^2}{k_x^2} \int dv_y dv_x \frac{ \paf{f_{\alpha 0}}{v_x}}{v_x - \omega/k_x - i\nu/k_x},
\end{align}
where $\omega_{ps}$ is the plasma frequency of species $s$, and $\nu \rightarrow 0^{+}$ determines the pole convention.
We further take $D_{r,\alpha} = 0$, which is a good approximation when the alpha particles are hot and sparse compared to the ions.

Plugging these dispersion components into the real dispersion Eq.~(\ref{eq:realDisp}), we find the familiar LH wave:
\begin{align}
	\omega_r = \pm \omega_{LH} \equiv \pm  \lp \omega_{pi}^{-2} + |\Omega_e \Omega_i|^{-1} \rp^{-1/2} .
\end{align}
Taking $\omega_r > 0$, the imaginary dispersion yields:
\begin{align}
	\omega_i &= \frac{\pi}{2} S_{k_x} \frac{\omega_{p\alpha}^2}{\omega_{pi}^2} \frac{\omega_{LH}^3}{k_x^2} \int dv_y \pa{f_{\alpha0}}{v_x}|_{y=y_0,v_x=v_p}, \label{eq:1dLHomegaI}
\end{align}
where $S_{k_x} = \sgn(k_x)$ determines the direction of the phase velocity $v_p \equiv \omega_r/k_x$.

To recover alpha channeling, we need to transform the distribution function and derivatives from phase space coordinates $x^i \equiv (x,y,v_x,v_y)$ to gyrocenter-energy coordinates $X^i \equiv (X,Y,K,\theta)$:
\begin{alignat}{2}
	&X = x + \frac{v_y}{\Omega_\alpha} && Y = y - \frac{v_x}{\Omega_\alpha}\\
	&K = \frac{1}{2} m \lp v_x^2 + v_y^2 \rp \qquad &&\theta = \arctan \lp -v_y, v_x \rp.
\end{alignat}
In this coordinate system, the phase space density function transforms as:
\begin{align}
	F_{\alpha 0} &= \sqrt{|g|} f_{\alpha 0}\\
	|g| &= |g_{ij}| = \left|\pa{x^m}{X^i} \pa{x^n}{X^j} \delta_{mn}\right| = m_\alpha^{-2}. \label{eq:metricTransformation}
\end{align}
Thus, we can rewrite our derivatives in terms of $X^i$ as:
\begin{align}
	\pa{f_{\alpha0}}{v_x}\biggr|_{y_0,\frac{\omega_r}{k_x}} &= \pa{X^i}{v_x} \pa{}{X^i}\lp m_\alpha F_{\alpha 0}\rp_{y_0,\frac{\omega_r}{k_x}}\\
	&=m_\alpha^2 \frac{\omega_r}{k_x}  \lp \pa{F_{\alpha0}}{K}  - \frac{k_x}{m_\alpha \omega_r \Omega_\alpha} \pa{F_{\alpha0}}{Y}\rp_{Y^*,K^*}. \label{eq:newDiffusionPath}
\end{align}
where $Y^* \equiv y_0-v_p/ \Omega_a, K^*\equiv  m \lp v_p^2 + v_y^2  \rp/2$, and we have taken $F_{\alpha 0}$ independent of $\theta$ to capture the gyrophase diffusion induced by the stochasticity.

The quantity in parentheses in Eq.~(\ref{eq:newDiffusionPath}) can be recognized as derivative along the diffusion path in Eq.~(\ref{eq:amplificationCondition}), and thus describes wave amplification from alpha channeling.
Interestingly, this means that the condition that there be (on average) a population inversion along the channeling diffusion path in gyrocenter-energy space is identical to the condition that there be a bump-on-tail instability in the local hot alpha particle distribution.

Under this new formalism, in contrast to the nonlinear formalism, it is possible to straightforwardly calculate the wave amplification rate.
For instance, for a Maxwellian with a gradient in $Y$, $F_{\alpha 0} = e^{-K/T_\alpha} e^{-Y/L} / 2\pi T_\alpha$, with thermal velocity $v_{th\alpha} = \sqrt{T_\alpha /m_\alpha}$, we have at $y=0$:
\begin{align}
	\omega_i &= -|\omega_{LH}|\sqrt{\frac{\pi}{8}}  \left| \frac{v_{px}}{v_{th\alpha}} \right|^3  e^{-\frac{1}{2} \lp \frac{v_{px}}{v_{th\alpha}}\rp^2} \notag\\
	&\qquad \times \left[ \frac{\omega_{p\alpha}^2}{\omega_{pi}^2} \lp 1  - \frac{v_{th\alpha}}{v_{px}} \frac{\rho_{th\alpha}}{L} \rp e^{-\lp y_0-\rho_{p\alpha}\rp/L}\right], \label{eq:channelingOmegaI1DMaxwellianLH}
\end{align}
where $\rho_{th\alpha} = v_{th\alpha}/\Omega_\alpha$, and $\rho_{p\alpha} = v_{p}/\Omega_\alpha$.

\begin{figure}[t]
	\center
	\includegraphics[width=0.7\linewidth]{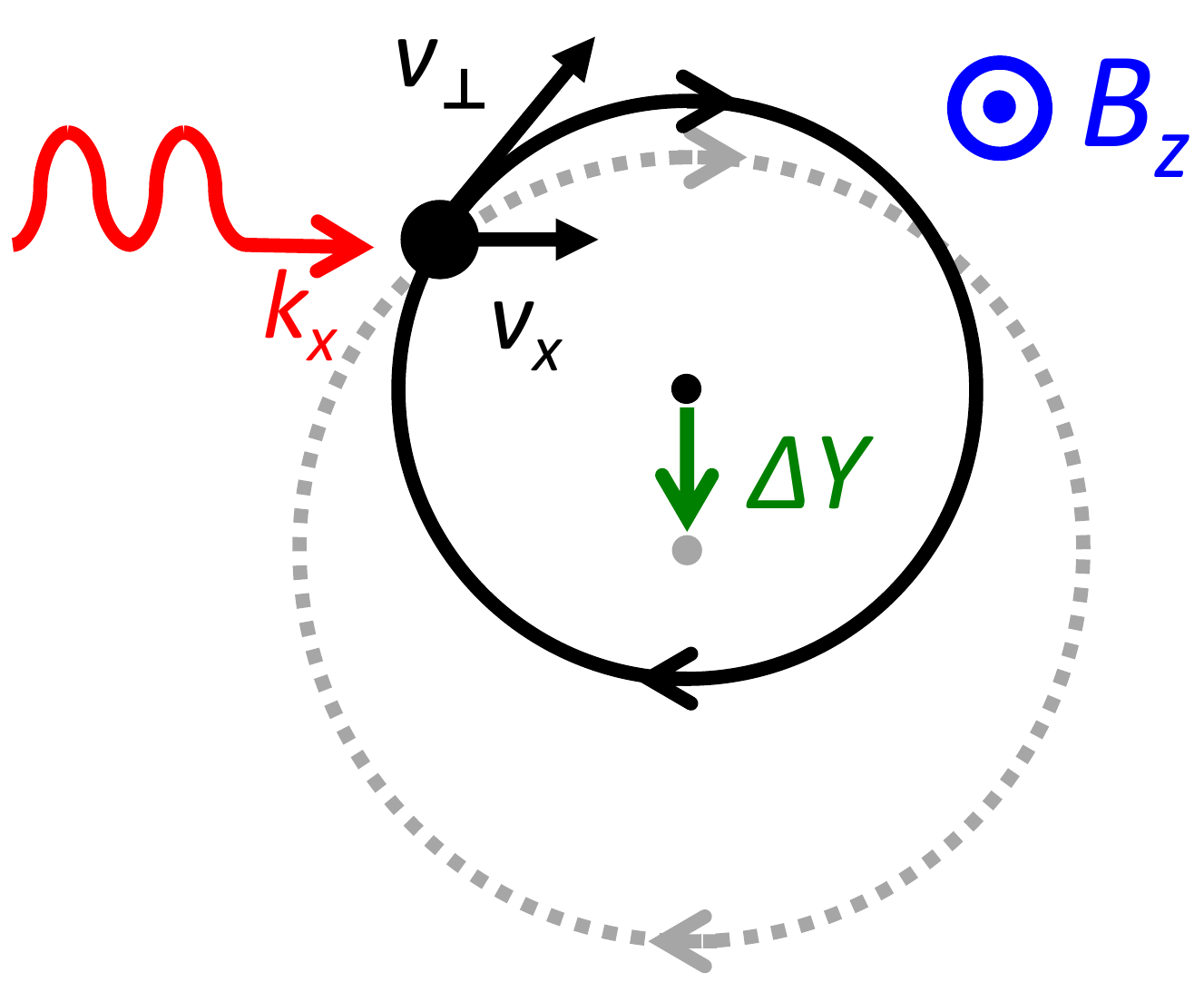}
	\caption{Schematic of the alpha channeling process.
		A hot particle with $v_\perp > v_p$ resonates with the wave at some point in the orbit, leading to a change in both energy and gyrocenter.
		Thus the particles diffuse along a specified path in gyrocenter-energy space.
		The amplification condition Eq.~(\ref{eq:amplificationCondition}) depends on the derivative of the distibution function along this path.
	}
	\label{fig:alphaChannelingSchem}
\end{figure}

\textbf{Resonant Diffusion:}
Identifying alpha channeling with the unmagnetized bump-on-tail instability allows us to compactly derive the diffusion tensor, by performing the same coordinate transformation.
From unmagnetized quasilinear theory, we have for a single wave mode \cite{kralltrivelpiece,Ochs2020}:
\begin{align}
	\pa{f}{t} &=  \pa{}{x_i}  \lp D_\ve{x}^{ij} \pa{f_{\alpha 0}}{v_j}\rp \label{eq:realSpaceDiffusion1dUnmagnetized}\\
	D_\ve{x}^{v_x v_x} &= \frac{\omega_{p \alpha}^2}{m_\alpha n_\alpha} \frac{W(y_0)}{k_x i} \lp \frac{1}{v_x - \omega_r/k_x - i\omega_i/k_x - i\epsilon} - c.c. \rp ,
\end{align}
where $W(y_0) = E_0(y_0)^2/16\pi$ is the wave electrostatic energy at $y_0$.
The diffusion equation~(\ref{eq:realSpaceDiffusion1dUnmagnetized}) transforms as:
\begin{align}
	\pa{F_{\alpha 0}}{t} &=  \pa{}{X^i}  \lp \sqrt{|g|} D_\ve{X}^{ij} \pa{}{X^j} \lp \frac{F_{\alpha 0}}{\sqrt{|g|}} \rp \rp, \label{eq:diffusionEqTransform}
\end{align}
where $D_\ve{X}^{ij}$ is determined from $D_\ve{x}^{ij}$ by the same tensor transformation law as for the metric in Eq.~(\ref{eq:metricTransformation}).

Performing the coordinate transformation, taking $F_{\alpha 0}$ independent of $\theta$, and averaging Eq.~(\ref{eq:diffusionEqTransform}) over $\theta$, we find:
\begin{align}
	\llangle \pa{F_{\alpha 0}}{t} \rrangle_t &= \pa{}{\bar{X}^i}  \lp  \llangle D_\ve{\bar{X}}^{ij} \rrangle_\theta \pa{F_{\alpha 0}}{\bar{X}^j}\rp,
\end{align}
where the gyro-averaged coordinates $\bar{X} \equiv (K,Y)$, and:
\begin{align}
	\llangle D_\ve{\bar{X}}^{ij} \rrangle_\theta &= \frac{1}{2\pi} \frac{\omega_{p\alpha}^2}{m_\alpha n_\alpha} \frac{W(y_0)}{k v_\perp i} \bvec 
	\Omega_s^{-2}  I_d^0& -\frac{m_\alpha v_\perp}{\Omega_\alpha} I_d^1  \\
	-\frac{m_\alpha v_\perp}{\Omega_\alpha} I_d^1 & m_\alpha^2 v_\perp^2 I_d^2 \evec  \label{eq:avgGyroDiffusion1}
\end{align}
with $v_\perp = \sqrt{2K/m_s}$, and $I_d^a = I_-^a - I_+^a$, where:
\begin{align}
	I^a_\pm &\approx \lp 1 \mp i \omega_i \pa{}{\omega_r} \rp \int_0^{2\pi} d\theta \frac{\sin^a \theta}{\sin \theta - \omega_r/k v_\perp \pm i \epsilon}.
\end{align}
This integral can be evaluated with the $u$-substitution $u = \sin \theta$.
We will focus on the resonant diffusion by ignoring for now the nonresonant contribution from $\omega_i$, which gives at $Y_0 = y_0 - v_p/ \Omega_\alpha$:
\begin{align}
	\llangle \pa{F_{\alpha 0}}{t} \rrangle_t  &= \frac{d}{dK}\biggr|_\text{path} \left[D^{KK} \frac{d}{dK}\biggr|_\text{path} F_{\alpha 0} \right] \label{eq:ChannelingPathDiffusionEquation}\\
	D^{KK} &\equiv \frac{m_\alpha^2}{2} \lp \frac{q_\alpha E_0(y_0)}{m_\alpha} \rp^2  \frac{v_p^2}{\sqrt{k^2 v_\perp^2 - \omega_r^2}} H\lp v_\perp -  v_p \rp
	 \label{eq:finalQLResonantGyroDiffusion}\\
	\frac{d}{dK}\biggr|_\text{path} &\equiv \lp \pa{}{K} - \frac{k}{m_s \omega_r \Omega_\alpha} \pa{}{Y}\rp.
\end{align}
Eq.~(\ref{eq:finalQLResonantGyroDiffusion}) is the same as Karney's \cite{karney1979stochastic} diffusion coefficent in $v_\perp$ in the limit of large $k_x \rho$ as used in \cite{fisch1992current} (see supplemental material).
Furthermore, the diffusion is seen to occur along the diffusion path in Eq.~(\ref{eq:amplificationCondition}), confirming that this approach recovers alpha channeling.

The diffusion coefficient corrects the energy-space diffusion coefficient in Ref.~\cite{Fisch1992} and Ref.~\cite{ochs2015alpha}.
This discrepancy is discussed in the supplemental material.
This error did not affect the study of alpha channeling in toroidal geometry due to ion-Bernstein waves (IBWs) \cite{herrmann1998cooling,Clark2000}, which relied on a different diffusion coefficient from orbit-averaging the cyclotron-resonant response \cite{Kaufman1972,Eriksson1994}.

\begin{figure}[t]
	\center
	\includegraphics[width=\linewidth]{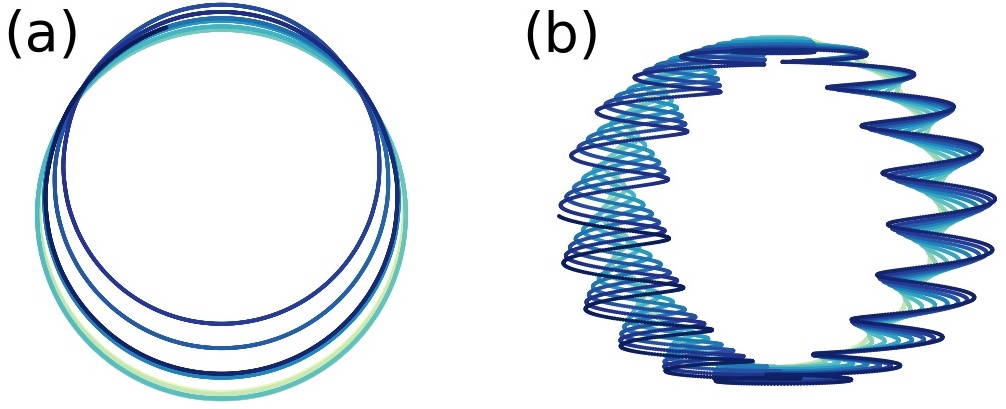}
	\caption{Simulated single-particle trajectories in the $x$-$y$ plane of (a) hot particles and (b) cold particles (relative to the phase velocity) in a growing electrostatic wave.
		The color of the trajectory indicates time, light to dark.
		The cold particles have a clearly non-gyrotropic velocity distribution due to the oscillations, and exhibit a clear shift in gyrocenter downward.
	}
	\label{fig:particleOrbits}
\end{figure}

\textbf{Nonresonant Reaction:}
Having established that the linear-quasilinear system recovers alpha channeling, we are now in a position to examine the nonresonant response.
In contrast to the resonant particles, which remain on largely unperturbed gyro-orbits except at the resonance points and thus have a largely gyrotropic distribution function (Fig.~\ref{fig:particleOrbits}a), the nonresonant particles experience sloshing motion along $v_x$ only, and thus have a nongyrotropic distribution function at $\mathcal{O}(E^2)$ (Fig.~\ref{fig:particleOrbits}b).

Thus, instead of transforming the nonresonant diffusion coefficient to the coordinates $X^i$, we find the nonresonant response by first calculating the total force density on species $s$ from the field-particle correlation:
\begin{align}
	F_{sx} &= q_s \langle E_{1x} n_1 \rangle_x. \label{eq:fieldParticleCorrelation}
\end{align}
This approach is equivalent to finding the force from the full (non-gyro-averaged) quasilinear theory.

Linearizing and Fourier transforming the above gives:
\begin{align}
	F_{s x} &= q_s \frac{1}{L} \int_{-L/2}^{L/2} dx \int \frac{dp_x'}{2\pi} \frac{dk_x'}{2 \pi} \tE_{p_x'} \tn_{s,k_x'} \notag\\
& \qquad \qquad \qquad \qquad \times	e^{i (p_x'+k_x')x - i(\omega(p_x') + \omega(k_x')) t}, \label{eq:fpcFourier}
\end{align} 
which can be expressed entirely in terms of $\tphi$ by using $\tE_{p_x'} = -i p_x' \tphi_{p_x'}$ and $\tn_{s,k_x'}$ from Eq.~(\ref{eq:ESDispersionGeneral}).
Then, in this Fourier convention, the wave $\phi = \phi_0 \cos (k_x x - \omega t) e^{\omega_i t}$ corresponds to:
\begin{align}
	\tphi_{k_x'} = \pi \phi_0 \left[\delta (k_x'-k_x) + \delta (k_x'+ k_x)\right]. \label{eq:phikSingleMode}
\end{align}
Plugging this in to Eq.~(\ref{eq:fpcFourier}) and making use of the symmetry property $D(-k_x) = D^*(k_x)$ allows us to calculate the total force on species $s$ as:
\begin{align}
	F_{s x} &= \frac{\phi_0^2 e^{2\omega_i t}}{8\pi} \Im \left[ k_x k^2 D_s \right]\\& \approx  2 W k_x \left[ D_{is} +  \omega_i \pa{ D_{rs}}{\omega_r} \right]. \label{eq:fpcForceFinal}
\end{align}
Here, the first term on the RHS is the force density on the resonant particles, and the second term is the force density on the nonresonant particles.
This derivation generalizes the result for an unmagnetized plasma in Ref.~\cite{Ochs2020} to any electrostatic wave.
Summing over all species, we recover the imaginary component of the dispersion function, Eq.~(\ref{eq:imagDisp}).
Thus the total force applied to the plasma sums to 0, as demanded by momentum conservation for the electrostatic wave.

The fact that the forces sum to zero in turn means that the total cross-field currents from resonant and non-resonant particles sum to zero, as can be seen by calculating the total current from the resulting $\ve{F} \times \ve{B}$ drifts:
\begin{align}
	\sum_s j_{sy} &= \sum_s -q_s n_s \frac{(F_{sx} / n_s) B_z}{q_s B_z^2} = -\frac{1}{B_z}\sum_s F_{sx} = 0.
\end{align}
Thus, for a purely poloidal wave mode growing in time, alpha channeling does not charge the plasma.

In addition to revealing the conservation of total charge, Eq.~(\ref{eq:fpcForceFinal}) also tells us which species carries the cancelling nonresonant current.
Specializing to an LH wave, with $\paf{D_{r,e}}{\omega_r} = 0$, we see that the nonresonant reaction is exclusively in the ions, which experience a force density:
\begin{align}
	F_{ix} &= 4 W k_x \omega_i \frac{\omega_{pi}^2}{\omega_r^3} = n_i m_i \lp \frac{q_i E_0}{m_i} \rp^2 \frac{k_x \omega_i}{\omega_r^3}. \label{eq:nonresIonForce}
\end{align}
Thus, for every alpha particle brought out of the plasma by the LH alpha channeling instability, two fuel ions are brought in, fueling the fusion reaction.

The total shift in the nonresonant ion gyrocenter due to the ponderomotive force for the LH wave can be expressed nicely by integrating the $\ve{F} \times \ve{B}$ drift over the growth of the wave, using $dE^2/dt = 2 \omega_i E^2$.
This gives:
\begin{align}
	\Delta Y &= -\frac{1}{2} \frac{q_i}{m_i} \lp \frac{k_x}{\omega_r} \rp^3 \frac{\Delta \phi_0^2}{B_z} . \label{eq:nonresGyroDisplacement}
\end{align}
In a multi-ion-species plasma, Eq.~(\ref{eq:nonresGyroDisplacement}) reveals to what extent each ion species moves inward as a result of LH alpha channeling.
For instance, in a p-B11 fusion plasma, Boron ions would pinch inward $Z_B/\mu_B = 5/11$ as much as the protons.
For other electrostatic waves such as the IBW, the general force density from Eq.~(\ref{eq:fpcForceFinal}) can be used to determine each species' response.
Eq.~(\ref{eq:nonresGyroDisplacement}) can also be easily checked against single-particle (Boris) simulations in which a wave is ramped up from 0 initial amplitude, which are found to agree well (Fig.~\ref{fig:gyrocenterShift}).
Details for these simulations are given in the supplemental material.

It is important to note that the cancellation of the resonant and nonresonant currents is not locally exact.
Because the nonresonant ions are cold, the electric field that enters this equation is evaluated locally at $y \approx Y$, in contrast to the case for the hot resonant particles, where it is evaluated at $y = Y+\omega_r/k_x \Omega_\alpha$.
Thus, if there is variation in the electric field in $y$ on some scale length $L$, the slight offset of the resonant and nonresonant currents will produce a net current ordered down from the resonant current by $\mathcal{O}(\rho_{p\alpha} / L) \ll 1$.
The resulting charge accumulation could in principle drive shear flow in the plasma, albeit at a much reduced rate than that suggested by the resonant current alone.

\begin{figure}[t]
	\center
	\includegraphics[width=\linewidth]{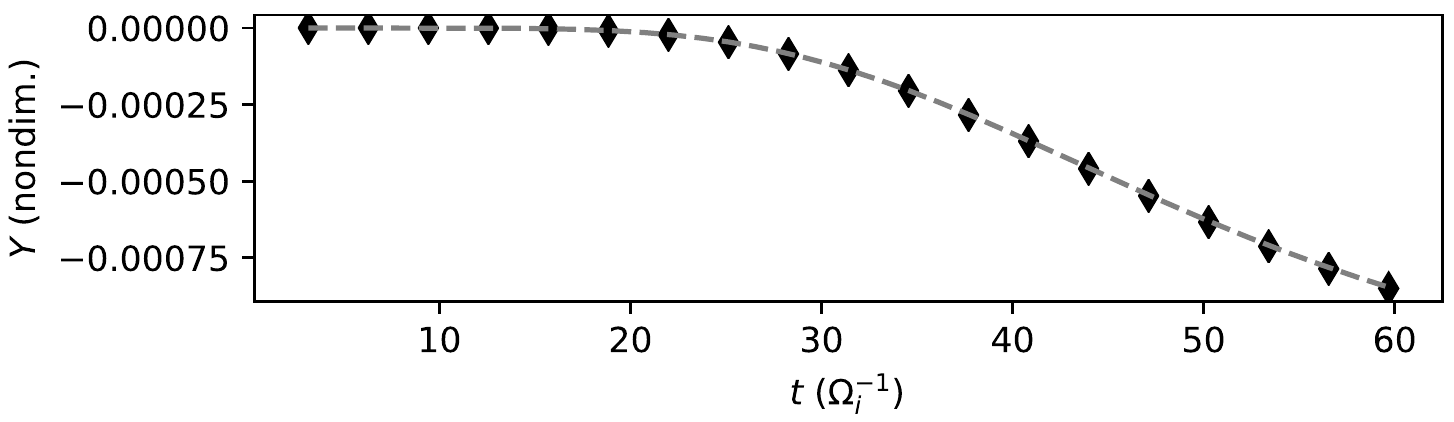}
	\caption{Change in gyrocenter position $Y$ for the particle in Fig.~\ref{fig:particleOrbits}b due to the slow ramp-up of the electrostatic wave.
	The gyro-period-averaged position in the simulations (filled black diamonds) agrees well with the predicted shift (gray dashed line) from Eq.~(\ref{eq:nonresGyroDisplacement}) due to the nonresonant reaction.
	}
	\label{fig:gyrocenterShift}
\end{figure}

\textbf{Discussion:} The force in Eq.~(\ref{eq:nonresIonForce}) is the same time-dependent force that arises from the unmagnetized ponderomotive potential in the form: 
\begin{align}
	\Phi = \frac{e^2 E_0^2}{4 m (\omega-\ve{k} \cdot \vv)^2}, \label{eq:unmagPonderomotive}
\end{align}
from whence the force is derived via:
\begin{align}
	(m \delta_{ij} - \Phi_{v_i v_j}) \frac{d v_j}{dt} = \Phi_{v_i x_j} v_j + \Phi_{v_i t} - \Phi_{x_i}.
\end{align}
where the subscripts represent derivatives. 
The force in Eq.~(\ref{eq:nonresIonForce}) appears as the second term on the RHS.

Eq.~(\ref{eq:unmagPonderomotive}) can be derived from \emph{unmagnetized} plasma susceptibility using the $K$-$\chi$ theorem \cite{Cary1977}, which relates the linear susceptibility to the ponderomotive potential.
Interestingly, application of the $K$-$\chi$ theorem to the \emph{magnetized} hot plasma dispersion relation \cite{stix1992waves} does \emph{not} yield the nonresonant force we observe here, as we show in the supplemental material.
Nevertheless, single-particle simulations using the Boris algorithm confirm the effect.
The failure of the hot plasma dispersion to capture the effect is likely related to the gyro-average in the hot plasma dispersion, which has been found to obscure the derivation of perpendicular resonant quasilinear forces \cite{Lee2012,guan2013toroidal,guan2013plasma}.

The approach used here, of taking the lowest-order alpha particle motion to be a straight trajectory, and then averaging over a gyroperiod, is similar to how neoclassical wave-particle interactions are treated. 
In those interactions, one does not generally use the full constant-of-motion space dispersion relation \cite{Kaufman1972} to calculate the quasilinear diffusion, but rather averages the effect of the diffusion derived from the magnetized dispersion relation over the neoclassical orbit \cite{Eriksson1994,Herrmann1997}.
This destroys resonances associated with the neoclassical orbit period, which are assumed to be destroyed by nonlinearities anyway.
In each case, the long-term orbit is ignored in the calculation of the dispersion, allowing in the neoclassical case cyclotron damping for banana orbits, and in the LH alpha channeling case Landau damping at resonance points on the gyro-orbit.

Note that, while the charge transport cancellation result is general for any purely poloidal electrostatic wave, the channeling path in Eq.~(\ref{eq:amplificationCondition}) and diffusion coefficient in Eq.~(\ref{eq:finalQLResonantGyroDiffusion}) apply only to the case of gyro-averaged Landau resonance \cite{Fisch1992,fisch1992current,ochs2015alpha,Cook2010}, and not to channeling via cyclotron resonances, as for the IBW \cite{Valeo1994,herrmann1998cooling,Clark2000,Cianfrani2019,Castaldo2019,Romanelli2020}.

\textbf{Conclusion:}
The alpha channeling interaction, which releases the free energy of  particles through diffusion in coupled energy-space coordinates, can rigorously be transformed to the classic bump-on-tail instability in velocity space only.  Applying the traditional mathematical apparatus then shows that, in an initial value problem, where resonant ions are ejected as the electrostatic wave grows at the expense of the ion energy, those same waves must pull in a return current of nonresonant ions so as to draw no current.  
This unexpected result is related to the cancellation of  resonant and nonresonant currents in the bump-on-tail instability \cite{kralltrivelpiece,Ochs2020}, except that these newly-found currents are perpendicular to the magnetic field, rather than parallel. 
We also calculated for the first time the contribution to the imaginary component of the dispersion relation due to alpha channeling, a useful quantity for ray tracing calculations.  

We not only prove rigorously the current cancellation, but we also determine the extent to which each species contributes to this cancellation.  
For LH waves, the nonresonant ions are pulled into the plasma core; thus, while no rotation is driven, the fusion reaction is beneficially fueled as ash is expelled.
In p-B11 reactor, we showed that protons are drawn in at twice the rate of Boron. 

While the nonresonant particles have been ignored in alpha channeling theory up to this point, our analysis shows that they can have important zeroth-order effects on the plasma dynamics.
However, the specific problem we considered here is only part of the story; in the most useful scenarios, channeling is driven by a stationary wave propagating radially inwards from an antenna at the boundary, requiring a fundamentally 2D analysis.
While the 2D problem is outside the scope of this Letter, the 1D self-consistent theory of alpha channeling laid out here provides a sound basis for examining the nonresonant response in more general scenarios.

\textbf{Acknowledgments:}
We would like to thank E.J. Kolmes and M.E. Mlodik for helpful discussions.
This work was supported by grants  DOE DE-SC0016072 and   DOE NNSA DE-NA0003871.
One author (IEO) also acknowledges the support of the DOE Computational Science Graduate Fellowship (DOE grant number DE-FG02-97ER25308).


%



\end{document}




\title{Supplementary Material: Nonresonant Diffusion in Alpha Channeling}

\author{Ian E. Ochs and Nathaniel J. Fisch}
\affiliation{Department of Astrophysical Sciences, Princeton University, Princeton, New Jersey 08540, USA}

\date{\today}

\maketitle
\tableofcontents

\section{Comparison of Resonant Perpendicular Diffusion Coefficients}

The perpendicular diffusion equation given by Karney \cite{karney1979stochastic} in the highly stochastic limit is [see \cite{karney1979stochastic} Eqs.~(71), (69), (3a-c)]:
\begin{align}
	\pa{f}{t} &= \frac{1}{v_\perp} \pa{}{v_\perp} v_\perp D^{v_\perp v_\perp} \pa{f}{v_\perp} \label{eq:KarneyDiffusionEq}\\
	D^{v_\perp v_\perp} &= \frac{\pi}{4} \frac{(E_0/B_0)^2 \omega^2 \Omega_i}{k_\perp^2 v_\perp^2} \left| H_\nu^{(1)} (r) \right|^2 \label{eq:KarneyDiffusionCoef}\\
	\nu &\equiv \omega/\Omega_i\\
	r&\equiv k_\perp v_\perp / \Omega_i.
\end{align}
It should be noted that this form of the diffusion equation treats arises by treating $f$ as a scalar function on phase space, so that probability integrals must include the metric of the transformation, i.e.
\begin{align}
	P(v_\perp - v_{\perp 0} < \Delta v_\perp) = \int_0^{2\pi} d\theta \int_{v_{\perp0} - \Delta v_\perp}^{v_{\perp0} + \Delta v_\perp} dv_\perp v_\perp f(v_\perp).
\end{align}
This is in contrast to our approach in the main text, where the metric is included in the distribution transform.

For large $r / \nu$, i.e. large $v_\perp$ relative to $v_p = \omega / k_\perp$, we have [see \cite{karney1979stochastic} Eq.~(68a)]:
\begin{align}
	H_\nu^{(1)} (r) &\approx \lp \frac{2}{\pi} \rp^{1/2} (r^2 - \nu^2)^{-1/4}\\
	&\approx \lp \frac{2}{\pi} \rp^{1/2} \Omega_i^{1/2} (k_\perp^2 v_\perp^2 - \omega^2)^{-1/4}.
\end{align}
Since Karney works in SI, with $\Omega_i = q_i B_0 / m_i$, plugging this approximation for $\nu$ into the diffusion coefficient yields:
\begin{align}
	D^{v_\perp v_\perp} = \frac{1}{2} \lp \frac{q_i E_0}{m_i} \rp^2 \frac{\lp \omega / k_\perp v_\perp\rp^2}{\lp k_\perp^2 v_\perp^2 - \omega^2\rp^{1/2}}.
\end{align}
This simplified form is used for the velocity-space coefficient in the later channeling literature \cite{fisch1992current}.

To transform this form to energy space, we note that the factors of the metric in the diffusion equation are already taken care of in the inclusion of the factors of $v_\perp$ in Eq.~(\ref{eq:KarneyDiffusionEq}).
Thus, all we need to do is transform the diffusion coefficient according to:
\begin{align}
	D^{KK} & = \lp \pa{K}{v_\perp} \rp^2 D^{v_\perp v_\perp}\\
	&= \lp m_i v_\perp \rp^2 D^{v_\perp v_\perp}\\
	&= \frac{m_i^2}{2} \lp \frac{q_i E_0}{m_i} \rp^2 \frac{\lp \omega / k_\perp \rp^2}{\lp k_\perp^2 v_\perp^2 - \omega^2\rp^{1/2}}.
\end{align}
This agrees with the form of the diffusion coefficient in the main text.

Ref.~\cite{Fisch1992} also employs an energy-space diffusion operator, which is used later in Ref.~\cite{ochs2015alpha}.
Focusing on energy-space, their diffusion equation was of the form:
\begin{align}
	\pa{F_\alpha}{\tau} &= \pa{}{\epsilon} D^{\epsilon \epsilon} \pa{F_\alpha}{\epsilon}\\
	D^{\epsilon \epsilon} &=(\epsilon_w)^a \frac{\omega}{\nu} \lp \frac{v_{osc}}{v_\alpha} \rp^2\frac{1}{(\epsilon - \epsilon_w)^{1/2}}
\end{align}
with $a = -3$, and:
\begin{align}
	\tau &\equiv 2 \nu t\\
	\epsilon &\equiv \frac{v_\perp^2}{v_\alpha^2}\\
	\epsilon_w &\equiv \lp \frac{\omega}{k_\perp v_\alpha} \rp^2\\
	v_{osc} &\equiv \frac{q_i E}{m_i \omega},
\end{align}
where $\nu$ is the collision frequency and $v_\alpha$ is the alpha particle birth energy.

In terms of $D^{\epsilon \epsilon}$, $D^{KK}$ is given by:
\begin{align}
	D^{KK} &= \lp \pa{\tau}{t} \rp \lp \pa{K}{\epsilon} \rp^2 D^{\epsilon \epsilon}\\
	&= \lp 2 \nu \rp \lp \frac{m_i v_\alpha^2}{2} \rp^2 D^{\epsilon \epsilon}.
\end{align}
Thus, plugging in all the definitions, we find:
\begin{align}
	D^{KK} &= \frac{m_i^2}{2} v_\alpha^4  \lp \frac{\omega}{k_\perp v_\alpha} \rp^{2a} \omega \lp \frac{q_i E}{m_i \omega v_\alpha} \rp^2 \frac{v_\alpha}{(v_\perp^2 - \omega^2/k_\perp^2)^{1/2}}\\
	&= v_\alpha^{3-2a} \frac{m_i^2}{2} \lp \frac{q_i E}{m_i} \rp^2  \frac{\lp \omega/k_\perp \rp^{2a-1}}{(k_\perp ^2v_\perp^2 - \omega^2)^{1/2}}
\end{align}
This agrees with the resonant diffusion coefficient in the main text if $a = 3/2$, \emph{not} if $a=-3$, as used in Refs.~\cite{Fisch1992,ochs2015alpha}.


\section{Particle Simulations}

The particle trajectories presented in the main text result from Boris \cite{boris1970relativistic} simulations of the Lorentz force, given in SI by:
\begin{align}
	m_s \frac{d \vv}{dt} &= q_s \lp  \Ev + \vv \times \Bv \rp.
\end{align}
We employ nondimensionalized units where $q_s = m_s = 1$.
Futhermore, we take $\Bv = 1 \hat{z}$, which normalizes $t$ to the gyrofrequency $\Omega_s$.

The electric field in the simulations is ramped up from 0, according to:
\begin{align}
	\Ev = \hat{x}k_x \phi_0 \sin \lp k_x x - \omega t \rp \tanh^4 \lp 2 t/\tau_{0} \rp.
\end{align}
For the simulations presented, we had $\tau_0 = 20\pi$, i.e. the field ramped up over the course of 10 gyroperiods.
We also had $k_x = 7.9$, $\omega = 20.12$, with a phase velocity $\omega / k_x = 2.55$, and an amplitude $\phi_0 = 0.2$.

For both simulations, the particles were initiated with $x_0 = \pi/k_x$, $y_0 = 0$.
For the simulation in Fig.~2a, we had $\vv_0 = (5,0)$, while for the simulation in Fig.~2b we had $\vv_0 = (0,0.01)$.
The specifics of these initial conditions were not fundamental to the results, and with the exception of the verification of the gyrocenter shift presented in Fig.~3, the simulations are mainly intended to show qualititative features of the orbits; hence the exclusion of the axis labels in Fig.~2.

In order to get a well-behaved simulation with an adaptive timestep, it was necessary to slightly modify the Boris algorithm.
This modification is necessary since the Boris algorithm's good features necessitate evaluating the electric field at a time point exactly midway through the velocity timestep, which does not occur with a blindly-applied adaptive timestep.
Since this implementation is nonstandard, we summarize it here.

The regular Boris algorithm is defined by \cite{Qin2013}:
\begin{align}
	\frac{\xv_{k+1} - \xv_k}{\Delta t} &= \vv_{k+1} \label{eq:BorisSpaceStep}\\
	\frac{\vv_{k+1} - \vv_k}{\Delta t} &= \frac{q_s}{m_s} \lp \Ev_k + \frac{(\vv_{k+1} + \vv_k ) \times \Bv_k}{2} \rp,
\end{align}
where $\xv = \xv(t_k)$, $\vv_k = \vv(t_k-\Delta t/2)$, $t_k = k \Delta t$, $\Ev_k = \Ev(\xv_k,t_k)$, and $\Bv_k = \Bv(\xv_k,t_k)$.

This procedure works when $\Delta t$ is constant; however, when $\Delta t = \Delta t_k$ is not constant, then the electric field will no longer be evaluated exactly between $v_k$ and $v_{k+1}$; instead, $\vv_k = \vv(t_k - \Delta t_{k-1}/2)$ and $\vv_{k+1} = \vv(t_k + \Delta t_k/2)$, while $\xv_k = \xv(t_k)$.

To get the adaptive timestep to work, we split the $\xv$ step in two.
This has the added benefit that output $(\xv_k,\vv_k)$ from the code represents $\xv$ and $\vv$ evaluated at the same timepoint $t_k$.
Thus, we define:
$\xv_k = \xv(t_k)$, $\vv_k = \vv(t_k)$.
Then, we have:
\begin{align}
	\frac{\xv_{k+1/2} - \xv_k}{\Delta t_k/2} &= \vv_{k}\\
	\frac{\vv_{k+1} - \vv_k}{\Delta t_k} &= \frac{q_s}{m_s} \lp \Ev_{k+1/2} + \frac{(\vv_{k+1} + \vv_k ) \times \Bv_{k+1/2}}{2} \rp\\
	\frac{\xv_{k+1} - \xv_{k+1/2}}{\Delta t_k/2} &= \vv_{k+1},
\end{align}
with $\Ev_{k+1/2} = \Ev(\xv_{k+1/2},t_k + \Delta t_k/2)$, and $\Bv_{k+1/2} = \Bv(\xv_{k+1/2},t_k + \Delta t_k/2)$.
Note that this procedure (a) ensures that the velocity timestep evaluates the fields at the exact midpoint time, and (b) implies the iteration equation for $\xv$:
\begin{align}
	\frac{\xv_{k+3/2} - \xv_{k+1/2}}{(\Delta t_k + \Delta t_{k+1})/2} &= \vv_{k+1},
\end{align}
which is the natural generalization of Eq.~(\ref{eq:BorisSpaceStep}) to an adaptive timestep, given the redefinition of $\xv_k$ to line up with $\vv_k$.
In effect, we have exchanged asymmetry in the $\vv$-step due to the adaptive timestep for asymmetry in the $\xv$-step, which in simple tests seems to have a less deleterious effect on the dynamics, allowing $\sim$10x larger timesteps than simply plugging a variable timestep into the Boris algorithm.

Our adaptive timestep itself was determined by:
\begin{align}
	\Delta t_k = \frac{1}{20} \min \lp \Omega_i^{-1}, \omega^{-1}, (k_x v_k)^{-1}, \sqrt{2 / k_x a_k} \rp,
\end{align}
with $a_k$ the acceleration; this meant that the timestep was always significantly less than that required for the particle to traverse the time or length scales of the problem.

\section{$K$-$\chi$ theorem for hot magnetized plasma}

The $K$-$\chi$ theorem \cite{Cary1977} relates the ponderomotive potential ($K$) to the dielectric susceptibility ($\chi$) of a plasma.
The $K$-$\chi$ theorem is written:
\begin{align}
	\int d\xv d\vv f_{s0} \Phi &= -\frac{1}{4\pi} \int d\xv \int d\xv' \Ev_\omega^*(\xv) \chi_\omega(\xv,\xv') \Ev_\omega (\xv'),
\end{align}
where
\begin{align}
	\ve{E}(\xv,t) = \Ev_\omega (\xv) e^{-i\omega t} + c.c.
\end{align}
Note that this relates to our variables via $|E_\omega|^2 = |E_0|^2/4$, yielding an extra factor of $1/4$, and uses a different Fourier convention (in terms of factors of $2\pi$) as well.
Furthermore, $f_s$ is here defined so that $\int d\vv f_s = n_s$; changing this normalization condition to equal 1 yields an extra factor of $n_s$ on the LHS.
Taking a homogenous plasma so that $\chi_\omega(\xv,\xv') = \chi_\omega(\xv-\xv')$, and considering a Fourier mode, we thus can write the $K$-$\chi$ theorem as:
\begin{align}
	n_s \int d\vv f_{s0} \Phi &= -\frac{E_0^2}{16\pi} \ve{e}_{\omega k}^* \chi_{\omega k} \ve{e}_{\omega k}, \label{eq:KchiFourier}
\end{align}
where $\ve{e}_{\omega k}$ is the normalized polarization vector.
From this form, the form in the main text quickly follows:
\begin{align}
	\Phi = -\frac{W}{n_s} \, \ve{e}^* \cdot \frac{\delta \chi_s}{\delta f_{s0}} \cdot \ve{e}. \label{eq:Kchi}
\end{align}

As an example, the Vlasov susceptibility for a hot, unmagnetized plasma is \cite{stix1992waves}:
\begin{align}
	\chi_s &= -\frac{\omega_{ps}^2}{k^2} \int d\vv \frac{k_i \paf{f_{s0}}{v_i}}{k_j v_j - \omega}\\
	&= -\int d\vv f_{s0} \frac{\omega_{ps}^2}{(k_m v_m - \omega)^2} \frac{k_i k_j}{k^2}.
\end{align}
Furthermore, for an electrostatic wave, the polarization vector is given by $e_i = k_i / k$.
Thus:
\begin{align}
	\Phi &= -\frac{W}{n_s} \, \frac{k_i}{k} \lp - \frac{\omega_{ps}^2}{(k_m v_m - \omega)^2} \frac{k_i k_j}{k^2} \rp \frac{k_j}{k}\\
	&= \frac{q_s^2 E_0^2}{4 m_s (\omega-\ve{k} \cdot \vv)^2}.
\end{align}

Now, we will apply the $K$-$\chi$ theorem to the hot magnetized susceptibility \cite{stix1992waves}, with $k_\parallel = 0$.
Since we are looking at electrostatic waves, we will have $\ve{e} = \hat{x}$, and we will only care about the $\chi_{xx}$ component of the susceptibility tensor.
We will examine the lowest-order thermal correction, where the time-dependent ponderomotive force makes its appearance in the unmagnetized case.

The hot magnetized susceptibility in this case is given by (suppressing species labels):
\begin{align}
	\chi_{xx} &= \frac{\omega_{p}^2}{\omega \Omega} \sum_{n=-\infty}^{\infty} \int_0^{\infty} 2\pi v_\perp dv_\perp \frac{\Omega}{\omega - n \Omega} S_{n,xx}\\
	S_{n,xx} &= \frac{n^2}{z^2} J_n(z)^2 v_\perp \pa{f_0}{v_\perp}\\
	z &\equiv \frac{k_\perp v_\perp}{\Omega}.
\end{align}
We can express this integral entirely in terms of $z$ by taking noting that $dv_\perp = (\Omega / k_\perp) dz$, and $\paf{f_0}{v_\perp} = (k_\perp / \Omega) \paf{f}{z}$, so plugging this in above and simplifying we find:
\begin{align}
	\chi_{xx} &= \lp \frac{\Omega}{k_\perp} \rp^2 \frac{\omega_{p}^2}{\omega \Omega} 2\pi \int_0^\infty dz \sum_{n=-\infty}^{\infty} \frac{n^2}{a-n} J_n(z)^2 \pa{f_0}{z},
\end{align}
where $a = \omega/\Omega$.

We can make use of the Newberger sum rule to derive the identity [see Ref.~\cite{swanson2012plasma} pg. 149]:
\begin{align}
	\sum_{n=-\infty}^{\infty} \frac{n^2}{a-n} J_n(z)^2 &= \frac{\pi a^2}{\sin \pi a} J_a(z) J_{-a} (z) - a.
\end{align}
Further expanding this in $k_\perp v_\perp / \omega = z/a \ll 1$, with $a>1$, we find:
\begin{align}
	\sum_{n=-\infty}^{\infty} \frac{n^2}{a-n} J_n(z)^2 &\approx \frac{a z^2}{2 (a^2 - 1)} + \frac{3}{8} \frac{a z^4}{a^4 - 5a^2 + 4} + \mathcal{O}\lp \frac{z}{a}\rp^6.
\end{align}

We plug this expansion in to our integral and integrate by parts to find:
\begin{align}
	\chi_{xx} &= \lp \frac{\Omega}{k_\perp} \rp^2 \frac{\omega_{p}^2}{\omega \Omega} 2\pi \int_0^\infty dz \left[-\frac{a z}{a^2 - 1} - \frac{3}{2} \frac{a z^3}{a^4 - 5a^2 + 4}\right]f_0.
\end{align}
Then, plugging back in for $a$ and $z$, we find:
\begin{align}
	\chi_{xx} &= \int_0^{\infty} \lp 2 \pi v_\perp dv_\perp \rp f_{0}  \notag\\
	&\qquad\times \biggl( -\frac{\omega_p^2}{\omega^2 -\Omega^2} - \frac{3}{2} \frac{\omega_p^2}{\omega^2} \frac{k_\perp^2 v_\perp^2}{\omega^2 - 5 \Omega^2 + 4 \Omega^4/\omega^2}\biggr) .
\end{align}
In this expression, the first term in brackets is the metric for the integral over phase space, after the integral over $\theta$.
Thus:
\begin{align}
	\frac{\delta \chi}{\delta f_0} &= -\frac{\omega_p^2}{\omega^2 -\Omega^2} - \frac{3}{2} \frac{\omega_p^2}{\omega^2} \frac{k_\perp^2 v_\perp^2}{\omega^2 - 5 \Omega^2 + 4 \Omega^4/\omega^2}.
\end{align}

Consider first the leading order term, $\Phi_0$.
Plugging in to the $K$-$\chi$ theorem Eq.~(\ref{eq:KchiFourier}), we have:
\begin{align}
	\Phi_0 &= \frac{q^2 E_0^2}{4 m (\omega^2 - \Omega^2)} .
\end{align}
This is the familiar ponderomotive potential for a cold plasma in the electrostatic limit \cite{Cary1977}.
It yields no $x$-directed force due to the growth of the field, since $\Phi_{0,v_x t} = 0$.

Next, we have the second-order term:
\begin{align}
	\Phi_2 &= \frac{q^2 E_0^2}{4 m  \lp \omega^2 - 5 \Omega^2 + 4 \Omega^4/\omega^2 \rp} \frac{k_\perp^2 v_\perp^2}{\omega^2} .
\end{align}
Since $v_\perp^2 = v_x^2 + v_y^2$, the $x$-directed time-dependent force density is given from this ponderomotive potential by:
\begin{align}
	F_x &= n \Phi_{v_x t}\\	
	&= \frac{q^2 n }{2 m  \lp \omega^2 - 5 \Omega^2 + 4 \Omega^4/\omega^2 \rp} \frac{k_\perp^2 v_x}{\omega^2} \frac{d E_0^2}{dt}\\
	&= \left[n m \lp \frac{q E_0}{m} \rp^2 \frac{k_\perp \omega_i}{\omega^3} \right] \frac{\omega k_\perp v_x}{\lp \omega^2 - 5 \Omega^2 + 4 \Omega^4/\omega^2 \rp},
\end{align}
where in the last line we used $dE_0^2 / dt = 2 \omega_i E_0^2$.
The first term in brackets here is the force density from the main paper, which was found to be consistent with the single-particle simulations.
Thus, the time-dependent ponderomotive force from the $K$-$\chi$ theorem is ordered down (in the relevant limit $\omega \gg \Omega$) by a factor $v_\perp / v_p \ll 1$ compared to the force in the simulations.
In short, this calculation fails to capture the force on nonresonant particles.

Likely, the failure of the $K$-$\chi$ theorem in this case comes from the fact that the susceptibility tensor is gyro-averaged prior to application of the theorem, thus losing the angyrotropy of the velocity distribution that gives rise to the nonresonant reaction.
Such gyrophase-dependent structure has been shown to also be important in evaluating perpendicular forces in resonant diffusion in magnetized plasmas \cite{Lee2012,guan2013toroidal,guan2013plasma}, so it makes sense that it could effect the ponderomotive forces as well.

%